\documentclass[conference]{IEEEtran}
\IEEEoverridecommandlockouts

\usepackage{cite}
\usepackage{amsmath,amssymb,amsfonts}
\usepackage{algorithm}
\usepackage{algorithmic}
\usepackage{graphicx}
\usepackage{multirow}
\usepackage{booktabs}
\usepackage{textcomp}
\usepackage{xcolor}
\def\BibTeX{{\rm B\kern-.05em{\sc i\kern-.025em b}\kern-.08em
    T\kern-.1667em\lower.7ex\hbox{E}\kern-.125emX}}

\usepackage{xcolor}
\newcommand{\Mozhgan}[1]{\textcolor{black}{#1}}
\newcommand{\Romina}[1]{\textcolor{black}{#1}}
\newcommand{\Task}[1]{\textcolor{black}{#1}}

\begin{document}

\title{EdgeNavMamba: Mamba‑Optimized Object Detection for Energy-Efficient
Edge Devices
}

\author{
\IEEEauthorblockN{Romina Aalishah, Mozhgan Navardi, and Tinoosh Mohsenin}
\IEEEauthorblockA{
Department of Electrical and Computer Engineering\\
Johns Hopkins University, Baltimore, MD, USA\\
}
}

\maketitle

\begin{abstract}

Deployment of efficient and accurate Deep Learning models has long been a challenge in autonomous navigation, particularly for real-time applications on resource-constrained edge devices. Edge devices are limited in computing power and memory, making model efficiency and compression essential. In this work, we propose EdgeNavMamba, a reinforcement learning-based framework for goal-directed navigation using an efficient Mamba object detection model.
To train and evaluate the detector, we introduce a custom shape detection dataset collected in diverse indoor settings, reflecting visual cues common in real-world navigation. The object detector serves as a pre-processing module, extracting bounding boxes (BBOX) from visual input, which are then passed to an RL policy to control goal-oriented navigation. Experimental results show that the student model achieved a reduction of 67\% in size, and up to 73\% in energy per inference on edge devices of NVIDIA Jetson Orin Nano and Raspberry Pi~5, while keeping the same performance as the teacher model. EdgeNavMamba also maintains high detection accuracy in MiniWorld and IsaacLab simulators while reducing parameters by 31\% compared to the baseline. 

\end{abstract}


\vspace{-3pt}

\section{Introduction}

\begin{figure}
    \centering
    \includegraphics[width=0.9\linewidth]{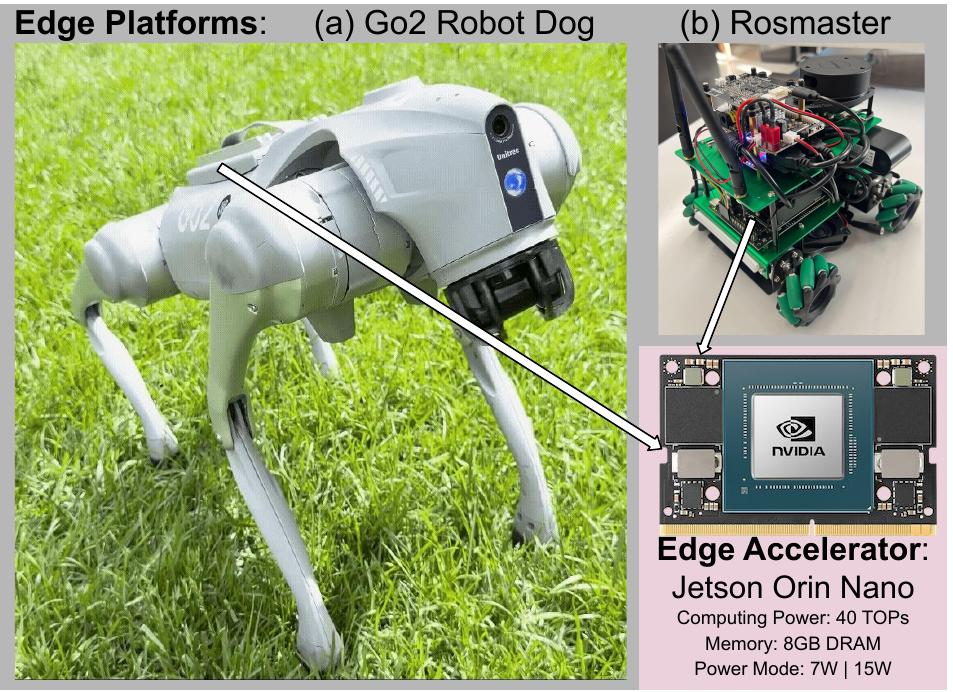}
    \caption{\Mozhgan{Edge platforms with onboard Jetson Orin Nano accelerators: (a) the Unitree Go2 robot dog and (b) the Yahboom Rosmaster wheeled robot.
    }}
    \label{fig:intro}
    \vspace{-16pt}
\end{figure}

Edge deployment is a key challenge for practical Deep Learning (DL) applications~\cite{wang2024computation, Navardi2025GenAI}, particularly in autonomous navigation, medical imaging , 
which require real-time performance~\cite{kallakuri2024atlas, walczak2025atlasv2, tahir2025edge, aalishah2025medmambalite, Xu2025EdgeDeepLearning, lee2024fast}. DL models on edge devices must be lightweight and efficient to provide real-time, reliable performance despite constraints in computation and power~\cite{aalishah2025mambalitesr, mozhgan2024metatinyml, mazumder2021survey}.  
\Romina{Particularly in autonomous navigation (Fig.~\ref{fig:intro}), scene understanding is critical, enabling vision models to learn environmental features, obstacles, and paths for navigation in both new and familiar scenarios~\cite{xie2025yoloace, walczak2025eden}. Deploying these models on edge devices is challenging due to their computational intensity, which is necessary for high accuracy
~\cite{mela2025yolo}.}

\Romina{Optimization methods
have been applied to these models to improve power and memory efficiency. Since You Only Look Once (YOLO)~\cite{redmon2016you} revolutionized object detection by 
using regression on bounding boxes, several efforts have applied these methods to YOLO. YOLO‐ACE redesigned the backbone and applied double distillation~\cite{xie2025yoloace}, 
and Mamba YOLO~\cite{Wang2025MambaYOLO} integrated a state‐space‐model~(SSM)~\cite{mamba} backbone for efficiency.
With the introduction of these lightweight yet powerful models, the deployment of edge devices for navigation tasks becomes more feasible and efficient.
For the navigation phase, Reinforcement Learning (RL) has been a successful inspiration, as it allows the agent to learn through interactions and real-time feedback~\cite{wang2024autonomous}. \Task{However, to the best of our knowledge no existing work has attempted to combine Mamba, Knowledge Distillation~(KD), and an optimization strategy to produce a model small enough to fit into cache memory, 
 thereby 
improving 
time and 
energy efficiency.}
}




\begin{figure*}
    \centering
    \includegraphics[width=0.9\linewidth]{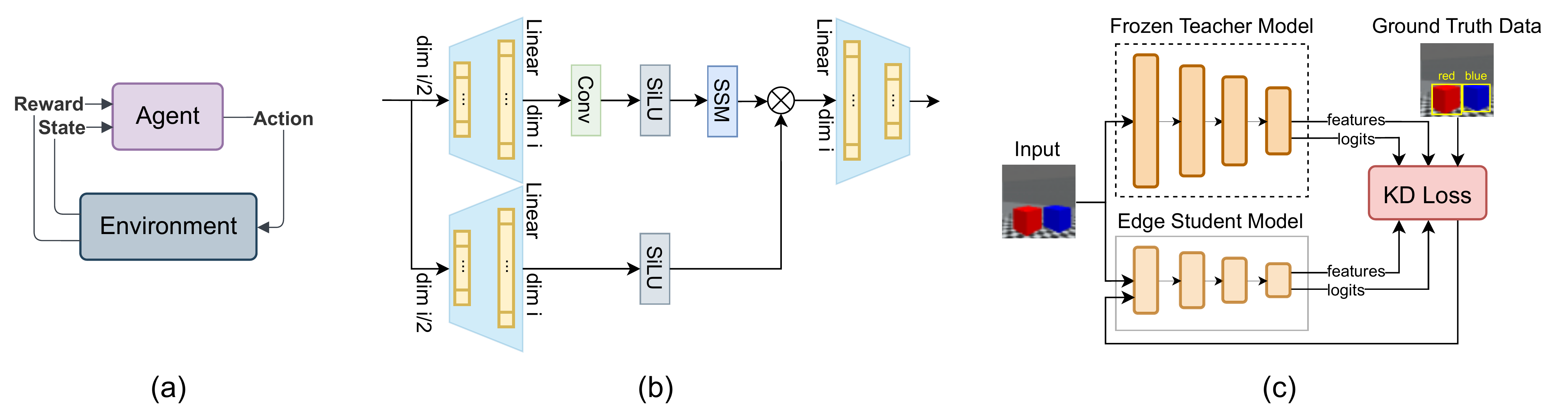}
     \vspace{-11pt}
    \caption{\Romina{(a) Reinforcement Learning (RL) diagram, including the interaction with the environment to maximize reward~\cite{sutton1998introduction, micro2023-mozhgan1}, (b) Architecture of Mamba~\cite{mamba}, used for feature extraction and model efficiency, (c) The process of knowledge distillation~\cite{hinton2015distilling}; the teacher model is trained and frozen, the student model is trained based on the teacher features, logits, and ground truth data.}}
    \label{fig:overview}
    \vspace{-11pt}
\end{figure*}

\Romina{
To address this,
\Task{we develop EdgeNavMamba, a customized Mamba-based detector tailored for efficient on-device perception. Unlike prior lightweight YOLO variants or state-space backbones, 
our design uniquely integrates the Mamba architecture with KD~\cite{hinton2015distilling} to achieve a balance between accuracy, 
and energy efficiency. The combination of state-space modeling and distillation enables compact yet context-aware feature representations that 
YOLO variants cannot capture. This framework directly addresses the memory and computational bottlenecks of edge deployment while maintaining real-time performance.}
We further validate the deployment 
of EdgeNavMamba on resource-constrained edge devices such as NVIDIA Jetson Orin Nano with 8 GB memory~\cite{nvidia_jetson_orin_nano} and Raspberry Pi 5 with 16 GB memory~\cite{raspberrypi}. The experimental results demonstrate that EdgeNavMamba successfully 
achieves efficiency with minimal performance loss compared to the teacher model.
}
\Romina{Our contributions are as follows:
\begin{itemize} 
    \item Development of an 
    edge Mamba object detector through architecture modification and knowledge distillation.
    \item Power and latency analysis for the proposed EdgeNavMamba on edge devices, such as the Raspberry Pi 5 and NVIDIA Jetson Orin Nano with Arm Cortex processors.
    \item Validation of object detection in simulators MiniWorld and IsaacLab, as well as RL navigation validation in MiniWorld with different complexities. 
\end{itemize}}

\Task{The rest of this paper reviews related work, outlines key preliminaries, introduces the EdgeNavMamba framework, 
and presents experimental results, and concludes with key findings.
}

\vspace{-5pt}

\section{Related Work}

Edge deployment is critical for real-world deep learning applications~\cite{wang2024computation, mozhgan-2025-AAAI-SSS-genai-}, especially in autonomous systems where onboard processing requires models to be light and efficient for real-time, reliable performance~\cite{tahir2025edge}.
Common optimization techniques include
architecture modification, knowledge distillation~\cite{hinton2015distilling}, quantization, and pruning~\cite{han2015deep, Navardi2025GenAI} 
Architecture 
changes adjust layer types, sizes, and their repetitions to maintain performance while reducing model size~\cite{aalishah2025medmambalite}. Knowledge distillation 
is applicable wether the teacher model is open-source or not~\cite{Navardi2025GenAI}. 
Quantization and pruning reduce memory usage by decreasing bit precision and removing connections in a structured or unstructured manner, respectively~\cite{han2015deep}.

\Romina{Object detection is one of the most computationally intensive tasks in computer vision and deep learning. Due to the need for high precision to detect objects of varying sizes, models are often large or require significant computational resources~\cite{mela2025yolo, humes2023squeezed}. Compression techniques address this issue. YOLO represents a major advancement in this field, addressing object detection in a regression-based manner~\cite{redmon2016you}. 
Lighter variants, such as 
YOLOv9~\cite{wang2024yolov9}, have been adapted for edge object deployment. To improve precision, newer versions
add an attention-based mechanism~\cite{tian2025yolov12},
but with a higher computational cost~\cite{mamba}. Mamba, a more efficient alternative to attention architecture, has been adopted in both full and hybrid forms in detection models~\cite{Wang2025MambaYOLO, visionmamba, mambavision}. With the introduction of these lightweight yet powerful models, the deployment of edge devices for navigation tasks becomes more feasible and efficient.
Mela et al. applied quantization and pruning for unmanned surface vehicles~\cite{mela2025yolo}.
Yang et al. 
proposed a multimodal 3D object detection framework using attention-guided and category-aware distillation~\cite{Yang2025MultiDistiller}.}

\Mozhgan{Reinforcement learning (RL) approaches such as deep Q‐networks (DQN)~\cite{mnih2015human} and proximal policy optimization (PPO)~\cite{schulman2017proximal} have been applied to autonomous navigation on resource‐constrained edge devices by directly mapping vision inputs to control commands~\cite{nahavandi2025comprehensive}. 
In~\cite{sim2real2022-mozhgan}, YOLO was integrated into a Deep Reinforcement Learning algorithm by passing the bounding‐box (BBOX) coordinates of \(n\) goals instead of raw images, improving training time and real‐world performance. However, as \(n\) grows, the input vector becomes larger, complicating goal learning, and adding a YOLO module adds significant edge-device overhead. To address this, a Squeezed‐Edge YOLO module integrated with RL was proposed to enhance the energy efficiency of the detection on edge devices~\cite{micro2023-mozhgan1, humes2023squeezed}.}

\Romina{In this work, we present an end-to-end framework for RL-based autonomous navigation with an optimized Mamba object detection model for energy-efficient edge computing. First, we design the optimized detector, which achieves competitive accuracy while using less memory and computation and therefore less energy than existing work. Next, we integrate this model into an RL algorithm and train the navigation policy in simulation. Finally, we deploy and evaluate the optimized object detection model on edge devices.}

\vspace{-5pt}

\section{Preliminaries}

\Romina{\textbf{Reinforcement Learning~(RL).} Goal-directed navigation, where the agent aims to reach an object in each episode, can be modeled as a Markov Decision Process (MDP)\cite{bellman1957markovian}, defined by a state space $S$, action space $A$, reward function $r: S \times A \rightarrow \mathbb{R}$, initial state distribution $s_0$, and transition probability $p(s_{t+1} \mid s_t, a_t)$.
RL\cite{sutton1998introduction} provides a set of algorithms that enable an agent to learn optimal policies $\pi(a \mid s)$ through trial-and-error interactions with the environment, aiming to maximize the cumulative expected reward. In goal-based tasks, the objective can be formulated as a goal-oriented MDP~\cite{reprohrl2023-tejaswini, micro2023-mozhgan1}, where RL methods learn to map states to actions that lead the agent toward the goal. Fig.~\ref{fig:overview}~(a) illustrates how an RL agent interacts with the environment under the MDP framework to receive rewards.}

\Romina{\textbf{Object Detection.} Object detection in computer vision aims to locate an object in images or videos by providing its spatial location in the form of bounding boxes and its category through class labels. The field is divided into two types of approaches: traditional techniques and machine learning-based methods. Traditional object detection methods rely on handcrafted features such as Haar~\cite{viola2001rapid}
, combined with brute-force techniques like sliding window searches across multiple scales and positions~\cite{viola2001rapid}. Due to their multi-stage pipelines, the introduction of YOLO as a real-time approach helped address it as a single regression problem~\cite{redmon2016you}. 
}

\Romina{\textbf{Mamba and State Space Models.} 
Mamba~\cite{mamba} architecture introduces Selective State Space Models, 
an efficient alternative to Transformers~\cite{he2016}, reducing computational complexity while maintaining feature extraction capabilities. The state space representation in Mamba is formulated as follows:}
\vspace{-10pt}

\begin{equation}
    \mathbf{y}(t) = \mathbf{C} \mathbf{x}(t) 
    \vspace{-1.5pt}
\end{equation}
\begin{equation}
    \frac{d}{dt} \mathbf{x}(t) = \mathbf{A} \mathbf{x}(t) + \mathbf{B} \mathbf{u}(t)
    \vspace{-1.5pt}
\end{equation}

\Romina{where $\mathbf{x}(t)$ represents the hidden state, $\mathbf{u}(t)$ is the input signal and $\mathbf{A}$, $\mathbf{B}$, and $\mathbf{C}$ are learnable matrices. This structure enables Mamba to capture long-range dependencies efficiently while requiring fewer parameters than traditional self-attention mechanisms. As a result, several efforts have been made to apply this method across various tasks. Fig.~\ref{fig:overview}~(b) demonstrates its architecture as a part of the network. 
}

\begin{figure}
    \centering
    \includegraphics[width=0.8\linewidth]{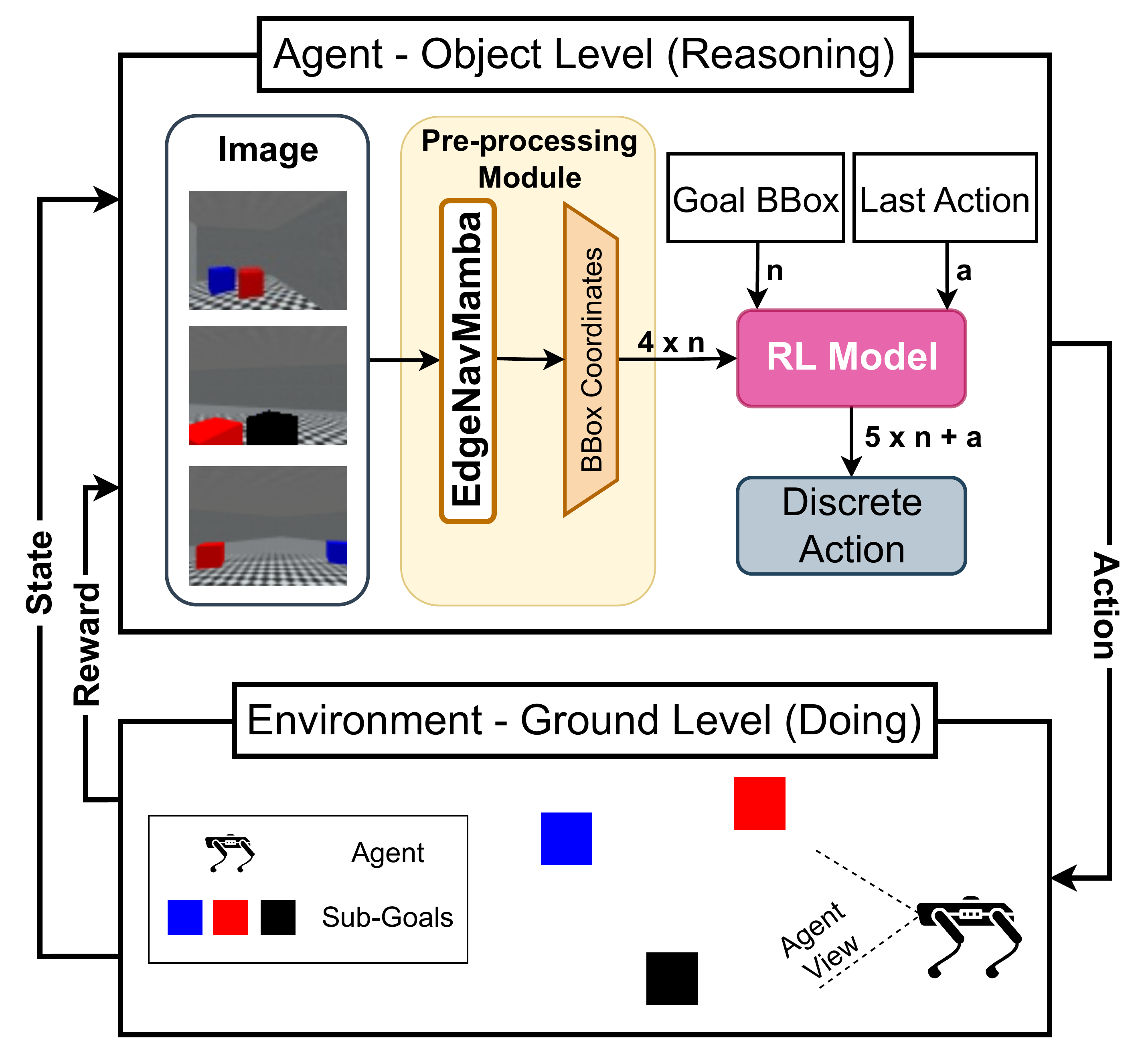}
    \caption{The proposed architecture of EdgeNavMamba, consisting of two branches of convolution and SSM for feature extraction. The same architecture 
    is used for teacher and student models 
    with a different set of dimensions. Features, then, undergo a detection process, and the bounding boxes are given to the RL model for navigation to the goal.}
    \label{fig:architecture}
    \vspace{-16pt}
\end{figure}

\Romina{\textbf{Knowledge Distillation~(KD).} Knowledge distillation transfers knowledge from a larger teacher model to a smaller student model to achieve similar performance with fewer parameters. 
In our setting, the student is trained using a combination of the standard YOLO detection loss, a distillation loss on classification logits, and a feature matching loss between intermediate representations:}
\vspace{-10pt}

\begin{equation}
L = L_{\mathrm{det}} + \lambda_{\mathrm{kd}} L_{\mathrm{KD}} + \lambda_{\mathrm{feat}} L_{\mathrm{feat}}
\label{eq:kd}
\vspace{-5pt}
\end{equation}


\Romina{where $L_{\mathrm{det}}$ is the standard YOLO detection loss computed from ground truth boxes and labels. $L_{\mathrm{KD}}$ is a temperature-scaled Kullback–Leibler divergence between the teacher and student classification logits controlled by a temperature parameter $T$. $L_{\mathrm{feat}}$ is the mean squared error between intermediate feature maps of the teacher and student.
The hyperparameters $\lambda_{\mathrm{kd}}$ and $\lambda_{\mathrm{feat}}$ control the relative contributions of the distillation and feature matching terms.}



\vspace{-5pt}

\section{Proposed Methodology}

\begin{figure}
    \centering
    \includegraphics[width=\linewidth]{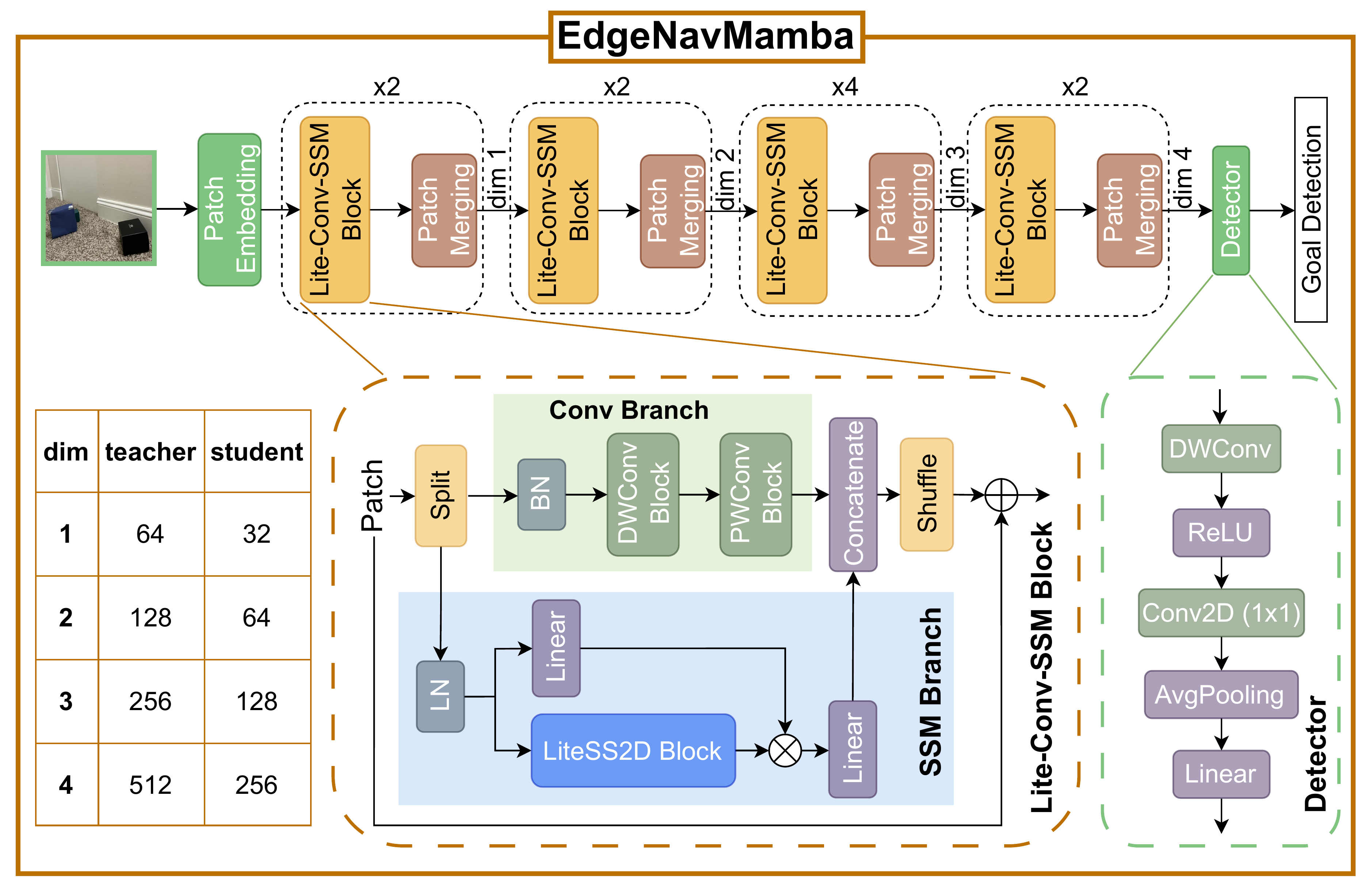}
    \caption{The proposed architecture of EdgeNavMamba, consisting of two branches of convolution and SSM (including LiteSS2D) for feature extraction. The same architecture 
    is used for teacher and student models 
    with a different set of dimensions. Features, then, undergo a detection process, and the bounding boxes are given to the RL model for navigation to the goal.}
    \label{fig:mambadet}
    \vspace{-16pt}
\end{figure}

\begin{figure*}
    \centering
    \includegraphics[width=0.9\linewidth]{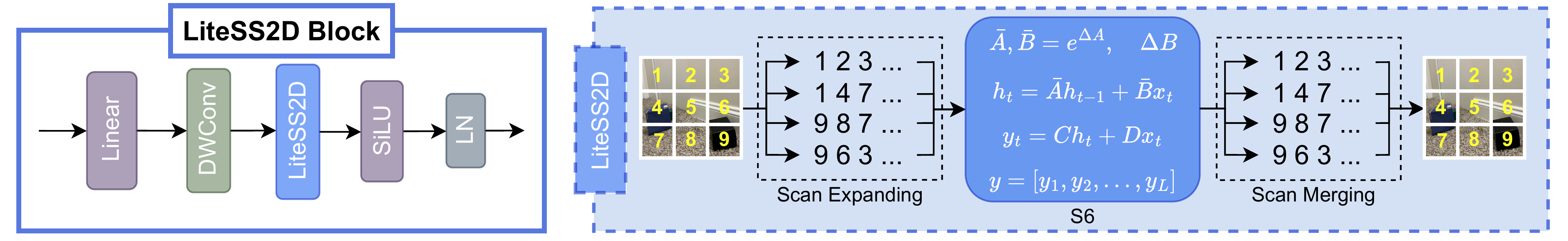}
    \caption{The architecture of LiteSS2D block, which is in the SSM branch of EdgeNavMamba, following the overall flow of the SS2D proposed by VMamba~\cite{liu2024vmamba}, but with modifications for better efficiency, mentioned in MedMambaLite~\cite{aalishah2025medmambalite}.}
    \label{fig:mamba}
    \vspace{-16pt}
\end{figure*}
\Mozhgan{In this section, we introduce the end-to-end framework called EdgeNavMamba for energy-efficient autonomous navigation, utilizing an optimized Mamba object detection model for on-device edge computing. Fig.~\ref{fig:architecture} and \Task{Algorithm~\ref{alg:edgenavmamba}} provide an overview of the proposed system. At each timestep, the agent captures an image of its environment, which the detector processes to extract BBOX coordinates of 
objects. These coordinates are then encoded as a feature vector and passed to an RL policy for goal navigation. Together, these components enable the agent to navigate autonomously to the goal while minimizing computations and energy usage. The RL policy is trained in MiniWorld and IsaacLab simulation environments. 
In the following section, we present our detailed approach.
}

\vspace{-4pt}

\subsection{Sim-to-Real Goal Navigation Framework}

\Romina{The navigation framework consists of two modules: an object detection network and an RL policy to reach the goal. First, the EdgeNavMamba 
processes the input image, divides it into a fixed resolution, 
and outputs normalized bounding boxes with confidence scores for $n$ detected objects. The resulting BBox coordinates $(x_1, y_1, x_2, y_2)$, producing a $1 \times (4n)$ vector. This is concatenated with a one-hot encoded sub-goal vector of size $1 \times n$, and a one-hot encoded last-action vector of size $1 \times a$, where $a$ is the number of discrete actions. resulting in a $1 \times (5n + a)$ state vector. During each episode in simulation, the PPO policy receives the full state vector, including all detected boxes plus one-hot goal, while reward shaping focuses on the BBox coordinates of the current goal object. The action space is discrete: $\{\texttt{left}, \texttt{right}, \texttt{forward}\}$. The PPO policy receives this state at each step and outputs an action. A task is considered complete when the agent comes within a predefined proximity threshold of the correct goal object, which checks the Euclidean distance between the agent and the target. 
Navigation is guided by the reward function shown in Table~\ref{tab:reward}. Distance change is to encourage the agent to reduce its distance to the goal at each step. First goal visibility and exploration rewards provide additional guidance when the goal is not yet in view, preventing the agent from remaining in states with no other positive reward signals.}

\begin{table}[t]
\centering
\caption{Reward components for navigation task in MiniWorld}
\label{tab:reward}
\begin{tabular}{ll}
\toprule
\textbf{Condition} & \textbf{Reward} \\
\midrule
Correct goal reached & $+10.0$ \\
Wrong goal / wall collision & $-2.0$ \\
Per step penalty & $-0.01$ \\
Opposite turn actions & $-0.05$ \\
Distance change & $0.5 \cdot (\Delta d_{\text{prev}} - \Delta d_{\text{curr}})$ \\
First goal visibility & $+0.1$ \\
Exploration (forward, no goal) & $+0.01$ \\
Exploration (turn, no goal) & $+0.005$ \\
\bottomrule
\end{tabular}
    \vspace{-13pt}
\end{table}

\subsection{Edge-Optimized Mamba Object Detection}
\textbf{Model Architecture.} Fig.~\ref{fig:mambadet} shows the overview of the proposed architecture for object detection, inspired by MedMambaLite~\cite{aalishah2025medmambalite}, which includes five main units as follows:

\subsubsection{Patch Embedding}

\Romina{The input image is split into patches and projected into a higher-dimensional space. }

\subsubsection{Lite-Conv-SSM-Block}
\Romina{Features pass through a series of Lite-Conv-SSM blocks, including convolutional and State-Space Modeling~(SSM) components. Convolutional branch captures local features using depthwise and pointwise convolutions. Meanwhile, the SSM branch utilizes a Lite 2D Selective Scan Mamba (LiteSS2D) module to capture long-range dependencies and global features. The outputs are concatenated and shuffled to fuse global and local features. A number of these blocks form stages in a hierarchical architecture.}

\subsubsection{Lite 2D-Selective-Scan}

\Romina{Fig.~\ref{fig:mamba} shows Lite 2D-Selective-Scan (LiteSS2D), which shares weights across four directions to reduce computation. The block starts by projecting the input features into a higher dimension, applies row-wise and column-wise convolutions, then runs a four-way Selective Scan with a shared SSM core.}

\Romina{\textbf{Scan Expanding:} flattens the input along four directions.}

\Romina{\textbf{S6 block:} processes each sequence with shared
weights.}

\Romina{\textbf{Scan Merging:} sums directional outputs and reshapes them.}

\Romina{This approach provides memory efficiency by avoiding repeated tensor reshaping and using compact representations. 
Compared to available object detection models, we introduced 
important changes to provide an 
efficient model. Efficiency is improved by factorizing convolutions, sharing projection weights, and reusing Mamba weight matrices across blocks. }

\subsubsection{Patch Merging}
\Romina{Between stages, patch merging layers reduce spatial resolution while increasing channel depth, building a hierarchical representation. }

\subsubsection{Detector}
\Romina{The detector processes extracted features to identify the presence and bounding box of a target object. It uses depthwise and pointwise convolutions, followed by pooling and a linear layer to output the
goal detection result.}

\textbf{Knowledge Distillation.} \Romina{Fig.~\ref{fig:overview} (c) illustrates our knowledge distillation framework, where an edge student model is trained based on a frozen teacher model and ground truth data. Models have a similar architecture, as shown in Fig.~\ref{fig:mambadet}, but with varying channel dimensions. During training, each input batch is processed by both teacher and student, and the student parameters are updated using a combined KD Loss according to the Eq.~\ref{eq:kd},
and optimization is performed on the student while keeping the teacher fixed.}



\begin{algorithm}[t]
\caption{EdgeNavMamba Proposed Approach
}

\label{alg:edgenavmamba}
\begin{algorithmic}[1]
\REQUIRE Dataset $\mathcal{D}$, teacher model $\mathcal{T}$, student model $\mathcal{S}$, RL policy $\pi$
\ENSURE Trained edge detector $\mathcal{S}^\star$ and navigation policy $\pi^\star$

\STATE \textbf{Train Teacher:} Train $\mathcal{T}$ on $\mathcal{D}$ using detection loss.
\STATE \textbf{Distill Student:} Freeze $\mathcal{T}$ and train $\mathcal{S}$ using 
$L = L_{\mathrm{det}} + \lambda_{\mathrm{kd}}L_{\mathrm{KD}} + \lambda_{\mathrm{feat}}L_{\mathrm{feat}}$.
\STATE \textbf{Train RL Policy:} Use $\mathcal{S}$ to extract object bounding boxes and feed them as state input to PPO agent $\pi$ in MiniWorld.
\STATE \textbf{Deploy on Edge:} Export $\mathcal{S}^\star$ and $\pi^\star$ to edge devices
for real-time goal navigation.

\end{algorithmic}

\end{algorithm}

\section{Experimental Evaluation}

\subsection{Experimental Setup}
\subsubsection{Datasets}
\Romina{
Two datasets were prepared for training and deployment \Task{of teacher and student models}: a real-world dataset containing 1,800 images and a simulated MiniWorld dataset with around 5,500 images. Both include three object classes, red, blue, and black boxes, and are split into training and validation sets with a 90/10 split.
}

\subsubsection{Training Details}
\Romina{For object detection model and knowledge distillation experiments, we set the temperature to $T=2.0$, the KL divergence weight to $\lambda_{\mathrm{kd}}=1.0$, and the feature-matching weight to $\lambda_{\mathrm{feat}}=0.25$. The teacher is first trained, then frozen, and distillation is performed into the student configured with depths $[2,2,4,2]$ and channel dimensions $(32,64,128,256)$ \Task{on the same dataset}.
We use the Adam optimizer with learning rate $\mathrm{lr}=10^{-4}$, batch size $32$, and a learning rate scheduler that reduces the rate when validation Mean Average Precision~(mAP) shows no improvement. Inputs are resized to $224\!\times\!224$ and normalized. Evaluation uses the mAP metric. During training and validation we periodically decode detections with confidence thresholds~$(0.25,\,0.45)$ for qualitative inspection. The trained student is exported to ONNX for deployment and integration into the RL network.}

\Romina{Navigation policy is trained in MiniWorld environment, consisting of a rectangular room with three colored boxes (red, blue, black) placed at random non-overlapping positions. At the beginning of each episode, one of the objects is randomly selected as the target, and its class is encoded as a one-hot goal vector. 
The policy is trained with the PPO algorithm. 
We use a learning rate of $3 \times 10^{-4}$, a batch size of $128$, and episode length of $1024$ steps.
The agent is trained for a total of $500{,}000$ timesteps.
The reward function is described in Table~\ref{tab:reward}, combining sparse success and failure signals with dense terms for distance reduction, exploration, and first-goal visibility. All experiments are done on an NVIDIA 4090 GPU.
}

\begin{figure}
    \centering
    \includegraphics[width=\linewidth]{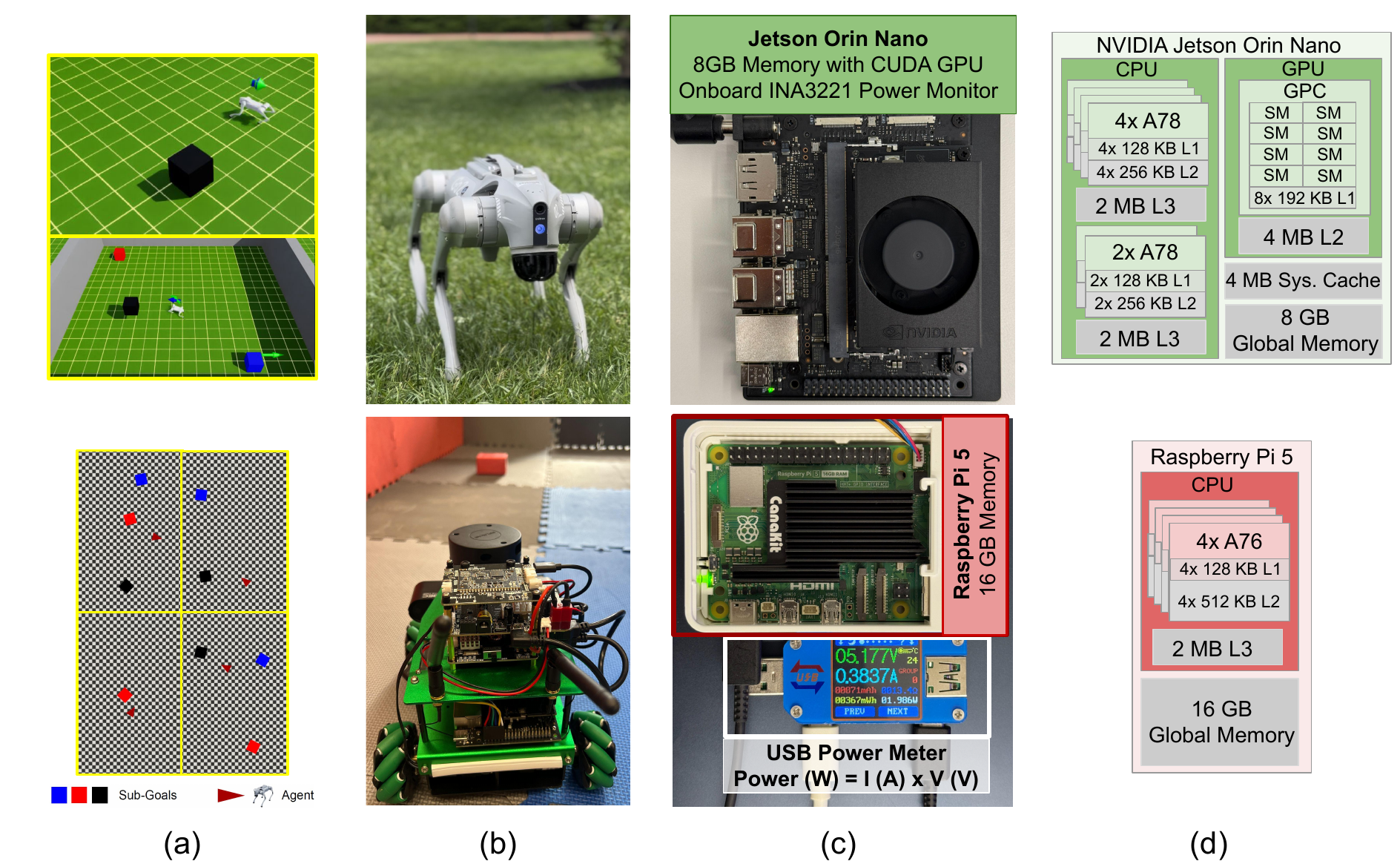}
    \caption{Experimental setup for energy-efficient multi-goal autonomous navigation: (a) simulation environment in NVIDIA Isaac Simulator and MiniWorld.
    \Mozhgan{(b) Edge robotic platforms: Yahboom Rosmaster wheeled robot and Unitree Go2 dog robot; (c) Raspberry Pi 5 edge node (4× Cortex‑A76 CPU, 16 GB LPDDR5, multi‑level cache); (d) NVIDIA Jetson Orin Nano 8 GB edge AI accelerator (6‑core Cortex‑A78AE CPU, Ampere GPU, 8 GB LPDDR5, multi‑level cache);
    (c) Cache hierarchy and power‑measurement setup using a USB power meter and onboard INA3221 monitor.}}
    \label{fig:exp_setup}
    \vspace{-12pt}
\end{figure}

\subsubsection{Hardware Deployment Platforms}
\Mozhgan{
To validate our approach in real-world settings, we deployed it on two edge platforms, an NVIDIA Jetson Orin Nano (8 GB RAM) and a Raspberry Pi 5 (16 GB RAM), mounted on the legged Unitree Go 2 robot and the Yahboom Rosmaster wheeled robot (Fig.~\ref{fig:exp_setup}~(b)). Power consumption was measured on both devices, as illustrated in Fig.~\ref{fig:exp_setup}~(c), and their memory hierarchies and CPU/GPU architectures are shown in Fig.~\ref{fig:exp_setup}~(d).}

\vspace{-7pt}
\subsection{Results and Discussion}
\subsubsection{Mamba Model Optimization}
\begin{figure}
    \centering
    \includegraphics[width=0.8\linewidth]{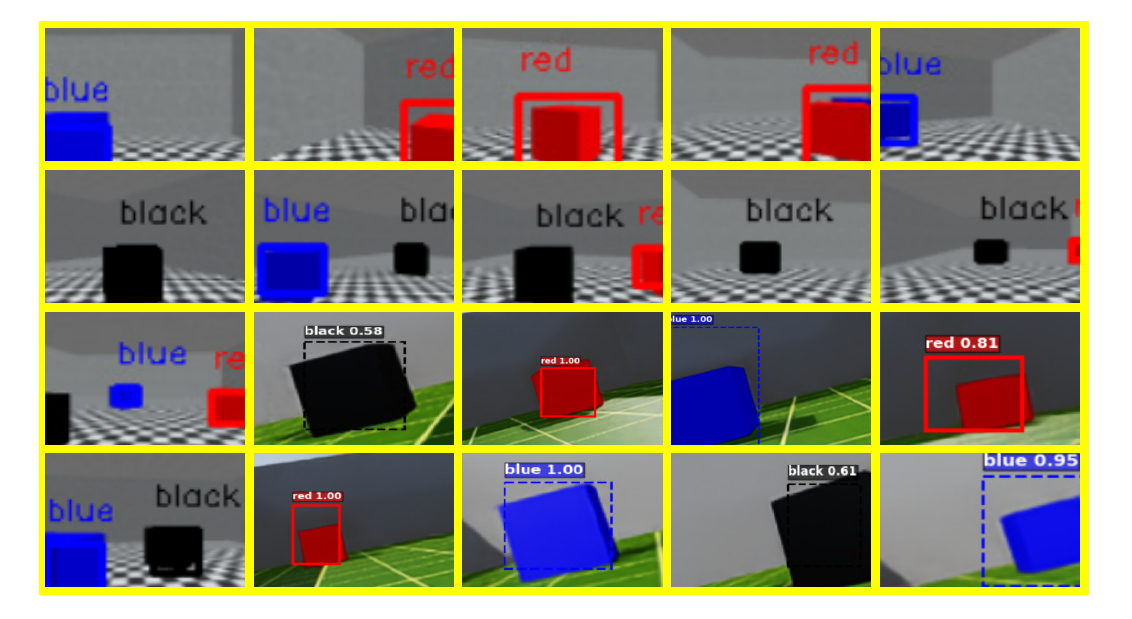}
    \caption{MiniWorld and IsaacLab samples of environments and object detections by the agent during exploration using EdgeNavMamba-ST. The environment contains three boxes placed at random, non-overlapping positions, with one randomly chosen as the target each episode.}
    \label{fig:miniworld}
    \vspace{-8pt}
\end{figure}

\Romina{Table~\ref{fig:comparison} presents the performance of the EdgeNavMamba teacher and student models in mAP compared to the existing
shape detection models. Knowledge distillation effectively reduces model size and FLOPs, without degrading performance. Meanwhile, our student model achieves a 31\% reduction in the number of parameters compared to the baseline, while maintaining competitive accuracy. Detections are evaluated in both MiniWorld and IsaacLab simulators for comprehensive analysis. Fig.~\ref{fig:miniworld} illustrates these environments along with examples of detections made by the agent in various scenarios.}

\subsubsection{RL‑Driven Goal Navigation}
\begin{figure}
    \centering
    \includegraphics[width=.9\linewidth]{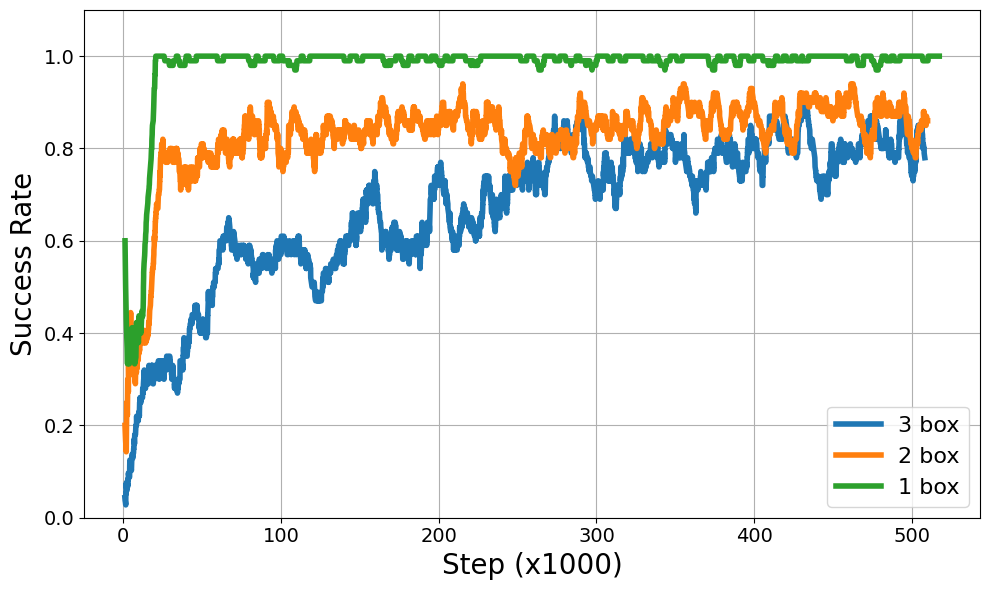}
    \caption{Success rate of navigation toward a defined goal during training in different environment complexities. Each value is calculated over the last 100 episodes. In each case, one box is designated as the goal, while the others serve as distractions.}
    \label{fig:success}
\end{figure}

\Romina{We evaluated EdgeNavMamba for navigation in MiniWorld using three scenarios. In each, one box was designated as the goal while the others served as distractions. In the first case, only one object was present; in the second, two objects were present, one being the goal; and in the third, three objects including one goal were placed. Fig.~\ref{fig:success} shows success rates for these scenarios over the last 100 training episodes. In the first case, the agent achieved a 100\% success rate, confirming accurate detection during navigation. In the second and third cases, the agent achieved 94\% and 90\% success rates, respectively.
}

\subsubsection{On‑Device Energy Profiling}

\begin{table}[t]
  \centering
  
  \caption{
  Comparison of EdgeNavMamba with prior models on the Shapes dataset. YOLOv5s~\cite{reprohrl2023-tejaswini,humes2023squeezed} (32-bit) and Squeezed Edge YOLO~\cite{humes2023squeezed} (8-bit) do not report FLOPs.
  }
  \vspace{-8pt}
  \label{tab:component_breakdown}
  \renewcommand{\arraystretch}{1.15}
  \setlength{\tabcolsep}{3pt}
  \begin{tabular}{lrrrr}
    \toprule
    \textbf{Block} &
      Params & Size & FLOPs & mAP \\
    \midrule
    \midrule
    \Task{\textbf{YOLOv5s~\cite{humes2023squeezed}}}       & \textbf{7.3 M} & \textbf{237 MB} & \textbf{-} & \textbf{0.96} \\
    \midrule
    \textbf{Squeezed Edge YOLO~\cite{humes2023squeezed}}       & \textbf{931 k} & \textbf{7.5 MB} & \textbf{-} & \textbf{0.95} \\
    \midrule
    \textbf{EdgeMambaNav-TR}       
    & \Task{\textbf{2.4 M}}
    & \textbf{9.1 MB} & \textbf{0.47 G} & \textbf{0.93} \\
    \midrule
    \textbf{EdgeMambaNav-ST}       & \textbf{639 k} & \textbf{2.5 MB} & \textbf{0.15 G} & \textbf{0.93} \\
    \midrule
  \end{tabular}
 \vspace{-8pt}
 \label{fig:comparison}
\end{table}

\begin{figure}[t]
    \centering
    \includegraphics[width=\linewidth]{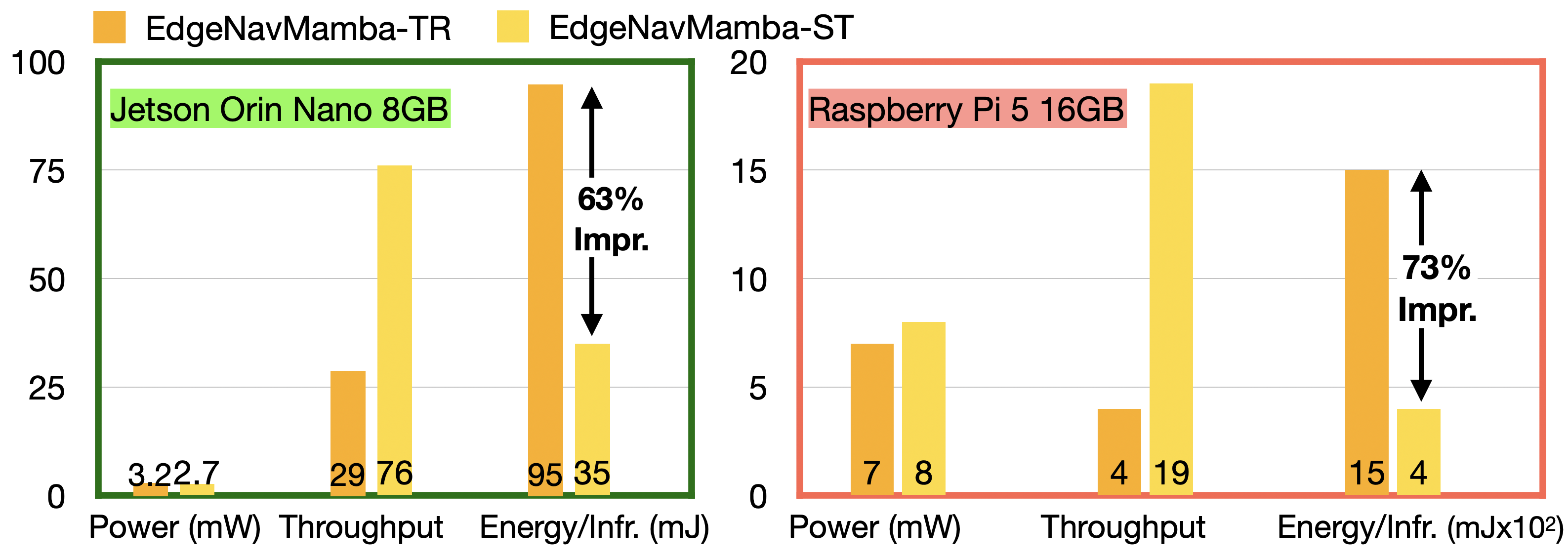}
    \caption{\Mozhgan{Energy and performance comparison of proposed EdgeNavMamba-TR and EdgeNavMamba-ST on Jetson Orin Nano~ and Raspberry Pi 5~16GB.}}
    \label{fig:hw_mamba_det}
\end{figure}

\Mozhgan{In Fig.~\ref{fig:hw_mamba_det}, we evaluate knowledge distillation by comparing the baseline EdgeNavMamba-TR with its distilled variant, EdgeNavMamba-ST, on two representative edge platforms. On the Jetson Orin Nano, EdgeNavMamba-ST achieves a 63\% reduction in energy per inference while improving throughput. Likewise, on the Raspberry Pi 5, EdgeNavMamba-ST delivers a 73\% energy reduction, demonstrating substantial efficiency gains with only negligible power overhead.}

\vspace{-5pt}

\section{Conclusion}
\Romina{In this work, we presented EdgeNavMamba, an RL-based framework designed for goal navigation using an efficient Mamba-based object detection model. By combining architectural modifications and knowledge distillation on the object detection model, we achieved a 31\% reduction in the number of parameters compared to the baselines while preserving detection accuracy. The student model also, achieved a reduction of 67\% in size, and up to 73\% in energy per inference on edge devices of NVIDIA Jetson Orin Nano and Raspberry Pi~5, while keeping the same performance as the teacher model, emphasizing the efficiency of the edge model. Navigation results in the MiniWorld simulator demonstrate over 90\% success rate in various environment complexities. 
}

\vspace{-5pt}

\bibliographystyle{IEEEtran}
\bibliography{ref.bib, eehpc.bib}

\vspace{12pt}

\end{document}